Lab-in-a-Tube: A portable imaging spectrophotometer for cost-effective, high-throughput, and label-free analysis of centrifugation processes

Yuanyuan Wei[1], Dehua Hu[1], Bijie Bai[2], Chenqi Meng[3], Tsz Kin Chan[1], Xing Zhao[4], Yuye Wang[5], Yi-Ping Ho[1, 6, 7, 8], Wu Yuan[1], Ho-Pui Ho[1, *]

[1] Department of Biomedical Engineering, The Chinese University of Hong Kong, Shatin, Hong Kong SAR, China. E-mail: aaron.ho@cuhk.edu.hk

[2] Electrical and Computer Engineering Department, University of California, Los Angeles, California, 90095, USA

[3] Department of Physics, The Chinese University of Hong Kong, Shatin, Hong Kong SAR, China

[4] Department of Orthopaedics and Traumatology, Li Ka Shing Institute of Health Sciences, The Chinese University of Hong Kong, Hong Kong SAR, China

[5] Bionic Sensing and Intelligence Center, Institute of Biomedical and Health Engineering, Shenzhen Institutes of Advanced Technology, Chinese Academy of Sciences, Shenzhen 518055, China.

[6] Centre for Biomaterials, The Chinese University of Hong Kong, Hong Kong SAR, China

[7] Hong Kong Branch of CAS Center for Excellence in Animal Evolution and Genetics, Hong Kong SAR, China

[8] The Ministry of Education Key Laboratory of Regeneration Medicine, Hong Kong SAR, China

[*]Correspondence: aaron.ho@cuhk.edu.hk

**Abstract**

Centrifuges serve as essential instruments in modern experimental sciences, facilitating a wide range of routine sample-processing tasks that necessitate material sedimentation. However, the study for real-time observation of the dynamical process during centrifugation has remained elusive. In this study, we developed an innovative Lab-in-a-Tube (LIAT) imaging spectrophotometer that incorporates capabilities of real-time image analysis and programmable interruption. This portable LIAT device measures 25.0 mm × 105.0 mm in size and weighs 52.8 g, fitting in common 50mL lab tubes. The cost is less than $30. Based on our knowledge, it is the first Wi-Fi camera built-in in common lab centrifuges with active closed-loop control. We tested our LIAT imaging spectrophotometer with solute-solvent interaction investigation obtained from lab centrifuges with quantitative data plotting in a real-time manner. Single re-circulating flow with a maximum fluid rotational velocity of up to 0.98 mm/s was real-time observed, forming the ring-shaped pattern during centrifugation. To the best of our knowledge, this is the very first observation of similar phenomena. Motivated by this observed phenomenon, we developed theoretical simulations for the single particle in a rotating reference frame, which correlated well with experimental results. We also demonstrated the field portability and on-site operation of our LIAT imaging spectrophotometer by performing an automated hematocrit test (Hct) of human whole blood from finger-prick volumes (20 μL). To the best of our knowledge, this marks the first demonstration to visualize the blood sedimentation process in clinical lab centrifuges. This remarkable cost-effectiveness opens up exciting opportunities for centrifugation microbiology research and paves the way for the creation of a network of computational imaging spectrometers at an affordable price for large-scale and continuous monitoring of centrifugal processes in general.

1. **Introduction**

Centrifuges are the workhorse in almost all medical diagnostics facilities. Centrifugation is a common technique in industry and scientific research for separating different constituent components inside a volume of liquid. It is also a crucial step in many diagnostic tests, assays, and therapeutic interventions ranging from the extraction of plasma from whole blood (for performing immunoassays or determining the hematocrit value) to the analysis of the concentration of pathogens and parasites in biological fluids, such as blood, urine, and stool (for microscopy) [1–4]. Despite its widespread use, applications of centrifugation have always been limited to material separation, and the current practice is both arbitrary and tedious, making a compromise in the outcome of the centrifugation step inevitable. The cost of this shortcoming might be innegligibly high if the constituent components to be extracted are otherwise difficult to produce and of extremely low quantity, not to mention that manual extraction may lead to unwanted band mixing and purity degradation for narrow bands. Besides, introducing automation in the centrifugation process is of great need for clinical scenarios. Taking blood component preparation (BCP) as an example, it saves a large number of repetitive manual actions[3]. Achieving automation of blood separation will certainly improve the efficiency of the reallocation and rescaling of the blood input, in order to relieve the stresses in blood banks and blood centers that are facing frequent urgent demands.

The main problem lies in the fact that currently all the standard centrifuge tubes are molded plastic containers, which means that they do not have any additional features for the "intelligent" processing of the sample. The design of commercially available centrifuge systems has remained almost unchanged over the past several decades – a rotating platform spinning the centrifuge tubes around[5]. The only advancements are integration of reagents within disposable cartridges. By changing centrifugal accelerations, entire assay protocol can be proceeded automatically. Even though it contains cost-efficient alternative for assay automation in laboratories with low to medium sample throughput, the progress within the centrifugation process is still invisible[6,7]. Referring to functional components and active control, a motor-assisted chip-in-a-tube (MACT)[5] has been developed into a 50 mL tube. The wirelessly controlled stepper motor induced one more directional control to manipulate the centrifugal force vector in a 3-dimensional (3D) manner. However, the MACT system still lacks the capability for real-time observation, leading to inevitable time and energy costs in repeated trials. The closest demonstrations to "real-time imaging of centrifugation process" are Lab-on-a-Disc (LOAD) systems or a wireless electrical resistance tomography (WERT) system. The LOAD systems usually consist of a disc, a rotor with an adaptor that fits the disc, and peripherals such as laser diodes, infrared lamps, magnets, and light-emitting diodes (LEDs)[8–10]. However, the cumbersome rotation system sacrificed their portability. In addition, the displacement scope of large-scale integration of bioassays are limited due to the limitation of disc radius[11]. Referring to the WERT system (diameter of D=95 mm, height of H=140 mm), it is capable for real-time imaging of particle distribution in centrifugal particles-liquid two-phase fields[12]. However, the WERT system has no adaption capability with common centrifuges. The rotational velocity of the WERT system is also limited to 1,490 r.p.m., far below the request of centrifugation processes requiring at least 3,000 r.p.m.. In conclusion, the real-time observation of the experimental process during centrifugation in common centrifuges has remained elusive.

The only major advancement comes from the development of analytical ultracentrifuges (for example, Beckman and Coulter's model OPTIMA AUC and model PROTEOMELAB XL series). They have optical windows along the centrifuge tube to allow for real-time optical supervision of the sample while it is undergoing centrifugation. Analytical information such as the spatially-resolved optical absorbance/fluorescence along the length of the sample column is obtainable from this technique, which can provide important properties of the solute components including molecular weight, diffusion coefficients, and solute-solvent interactions. Indeed, the emergence of analytical centrifuges has yielded a good amount of valuable data to help users understand various physical and chemical properties of the sample solute components[13–17]. However, the cost of this kind of facility is normally more than $300,000. A portable imaging platform for cost-effective, high-throughput, and label-free analysis of centrifugation processes adapting to common centrifuges remains essential and yet to be realized, especially for diagnostics in tiny-amount environments.

Here, to provide a powerful yet mobile and inexpensive tool for biochemical-related research, we introduce a wireless Lab-in-a-Tube (LIAT) imaging spectrophotometer system that can automatically detect and provide real-time color images of label-free objects inside a continuously centrifugated sample with a centrifugation speed of up to 3,000 r.p.m. The whole system weighs 52.8 g with size of a standard 50 mL centrifuge tube (diameter 25.0 mm × height 105.0 mm, see Fig. 1) and a cost of less than $30. Compared with other lab-on-a-disc systems or analytical ultracentrifuge, the presented device is significantly more compact, light-weighing, and extremely cost-effective. This device continuously examines processes within the microfluidic chip without any fluorescence triggering or hydrodynamic focusing of the sample, which also makes it robust and simple to operate, with a very large dynamic range in terms of the object size from microns to sub-millimeters. Integrated with a USB relay connecting the centrifuge plug and a laptop, the centrifugation can also be automatically suspended on demand. We demonstrated the real-time continuous observation and active control capabilities of our system by studying solute-solvent interactions and programmable interrupt of finger-prick volume(~20 μL) blood separation.

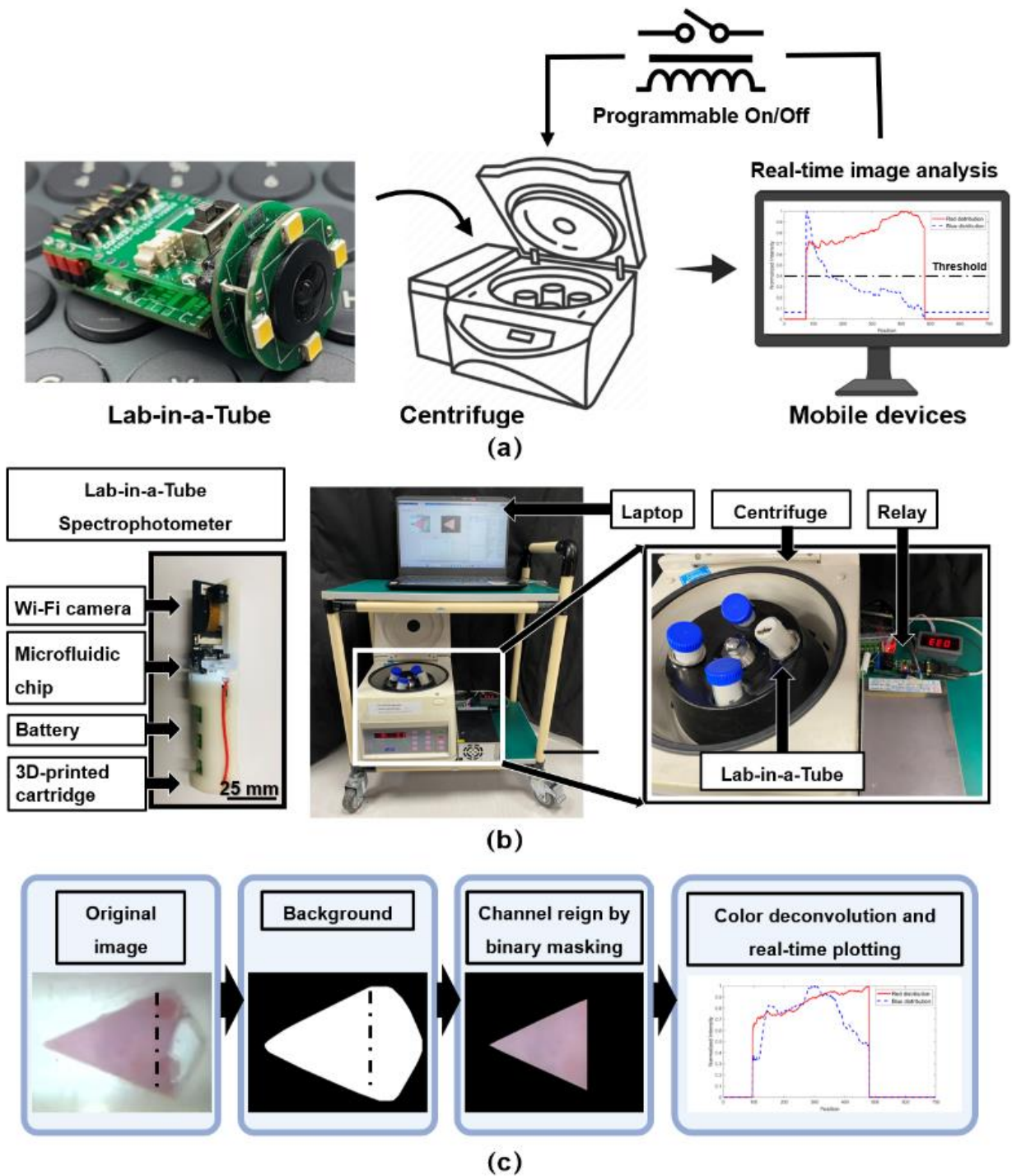

**Fig. 1 Construction and workflow of Lab-in-a-Tube (LIAT) imaging spectrophotometer system.** (a) Schematic of the LIAT spectrometer system. The captured images are real-time received by mobile devices outside the centrifuge while it is undergoing the centrifugation process. Integrated with color-based real-time image analysis, the centrifugation process could be programmable interrupt as soon as the detected intensity exceeds the set threshold. Based on this platform, sedimentation and mix have been observed and performed on demand in a real-time manner. (b) Construction of the LIAT spectrometer system including a Wi-Fi camera, replaceable

microfluidic chip, white light LED, lithium battery, and 3D-printed cartridge. This system is capable of real-time image acquisition and transmission to any mobile devices (including mobile phones and laptops) outside the centrifuge (SCEN-206 Digital Centrifuge, M.R.C. LTD. 0~6,000 r.p.m.), which makes it capable of monitoring the biological experiments inside the centrifuge and instantly controlling the centrifugal process. (Scale bar = 25 mm, the diameter of a 50-mL LabTube). (c) Workflow of color-based image analysis for real-time quantification. Channel region was segmented based on pre-experiment results. Based on the color deconvolution method, different colors can be plotted respectively.

## 2. Results

We tested our LIAT imaging spectrophotometer with solute-solvent interaction investigation obtained from lab centrifuges. Polystyrene beads (Sigma-Aldrich Inc., Korea) of different sizes and concentrations were injected into the fabricated microfluidic chips. The samples were in real-time imaged with a centrifugation velocity of up to 3,000 r.p.m. The raw full field of view (FOV) image information was saved on the controlling laptop. The Sample regions were segmented automatically by binary masking, and color intensities were quantified by the device using the color deconvolution method. The real-time display, real-time plotting, and programmable interruption of centrifugation were controlled through a custom-designed graphical user interface (GUI) (Supplementary Fig. 3). Figure 2 highlights the performance of this automated image analysis process and the function achieved by the device. As is shown in Fig. 2(a), sedimentation and separation process was continuously monitored and qualified by this device, showcasing different centrifugation process with both their initial raw captured images and the plotted intensity of different components. With the sedimentation process carried out, the peak of intensity moves to the left side, which correlates well with the captured frames. We were also able to observe the single re-circulating flow, or micro vortex, with a maximum fluid rotational velocity of up to 0.98 mm/s (radius = 1.63 mm). As shown in Fig. 2 (b), the core of the single re-circulating flow is shown to bend to the wall of the separator and rotate around the wall, forming the ring-shaped pattern normally observed at the vortex end. To the best of our knowledge, this is the very first observation of similar phenomena. To understand the dynamics inside the single re-circulating flow, we model the equilibrium motion of a single particle within the microfluidic chamber (in a rotating reference frame) shown in Fig. 2 (c). From simulation results, sedimentation and single re-criculating flow exist simultaneously, which correlates well with observed experimental phenomenon. Besides, the particle tends to rotate in the opposite direction of the rotating frame, and the centrifugal force dumps the particles away from the rotational axis. The particle rotates the rotational axis with the constant angular velocity but approaches the center of the axis exponentially. Details can be found in the Materials and Methods section.

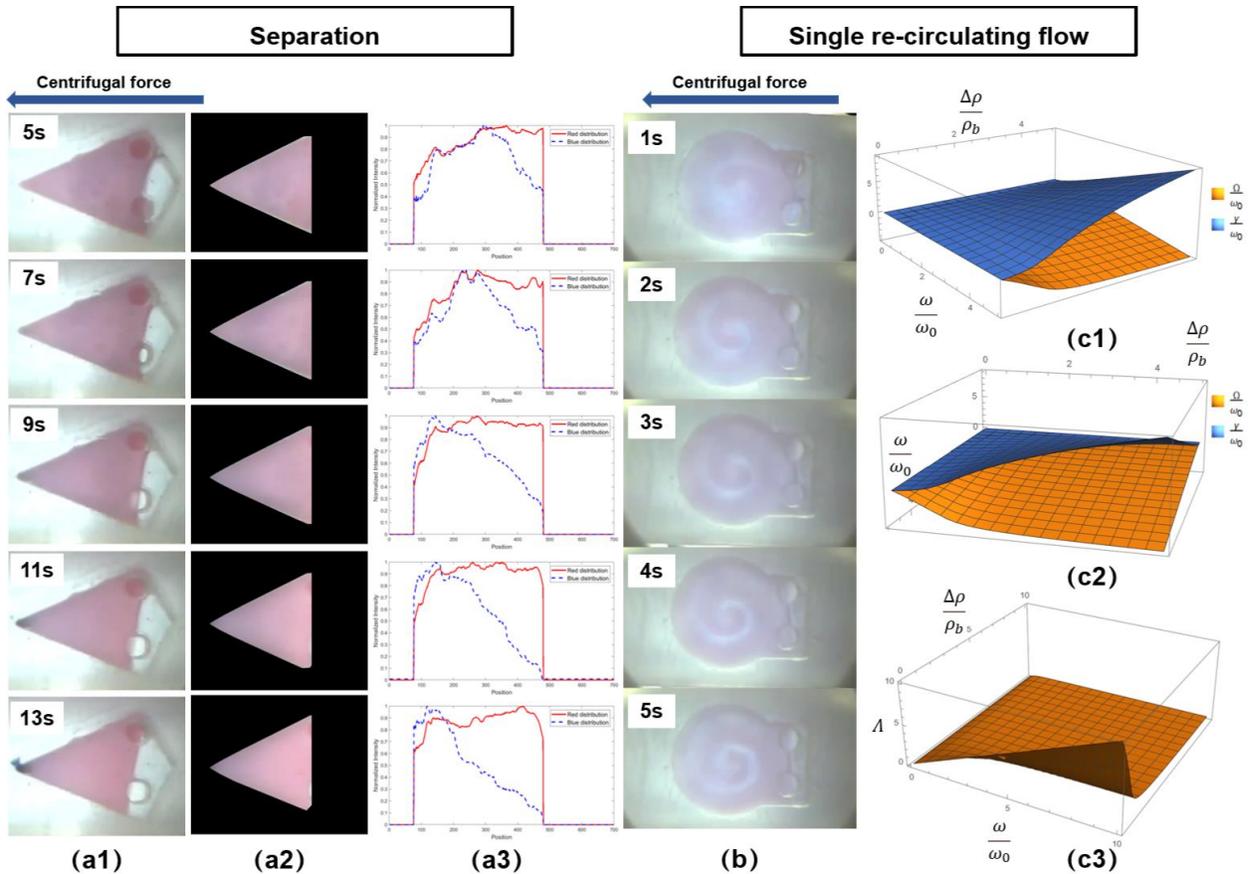

**Fig. 2 Real-time observation for sedimentation and single re-circulating flow of PS beads by LIAT imaging spectrophotometer system and simulation results.** Red beads sized 1 μm. Blue beads sized 10 μm. (a) Sedimentation and separation of different PS beads have been real-time observed and quantified. (a1) Representative frames of the centrifugation process with both 1 μm red beads and 10 μm blue beads injected into the microfluidic chip. (a2) Automatic segmentation result of sample region. (a3) Plotted distribution of red PS beads (1 μm) and blue PS beads (10 μm). The peak of blue PS beads shifted left side obviously. The distribution of red PS beads had no significant change. (b) Single re-circulating flow in centrifugal microfluidics has been observed for the first time. Representative frames of the centrifugation process with 1 μm red PS beads injected into the microfluidic chip. (c) Theoretical model for a single particle in the rotating reference frame. (c1) (c2) Angular velocity $\Omega < 0$ indicates the particle tends to rotate in the opposite direction to the rotating frame. Centrifugal relative velocity $\gamma > 0$ indicates the particle moving away from the rotational center axis. With radius and rotational angular velocity increase, the absolute value of $\Omega$ and $\gamma$ increase monotonically. This conclusion has also been proven by experiments: sedimentation and vortex exist simultaneously. (c3) The significance of the single-recirculating flow as the function of the parameters. The differentiation of orbiting angular velocities $\Lambda$ increases monotonically with the parameter $\omega/\omega_0$ but decrease with $\Delta\rho/\rho_b$. And $\Lambda$ increases particularly fast when the density difference is far more smaller than the density of the particle.

We tested our LIAT imaging spectrophotometer by monitoring and analyzing the centrifugal microfluidic sedimentation process of polystyrene beads. The PS beads of different sizes and concentrations were tested with the aforementioned method. Figure 3 (a) highlights the sequentially captured frames and correlated analysis results by plotting color intensity distribution. Detailed processes including the color calibration process can be found in the Discussion section. Here we define the sedimentation by an integral calculation of an "intensity function" over space (x-axis) as shown in Fig. 3 (b). In turn, the sedimentation speed correlates to the position moving speed of 90% normalized intensity function. The different between experimental results and calculated results (details in Materials and Methods section) may come from sample collection, imaging, and data processing techniques. However, our method is self-consistent, and therefore the relative differences are comparable, illustrating the real-time monitoring and analysis capability of our device. Results of other experimental conditions can be found in Supplementary Information (Supplementary Fig. 6 and Fig. 7).

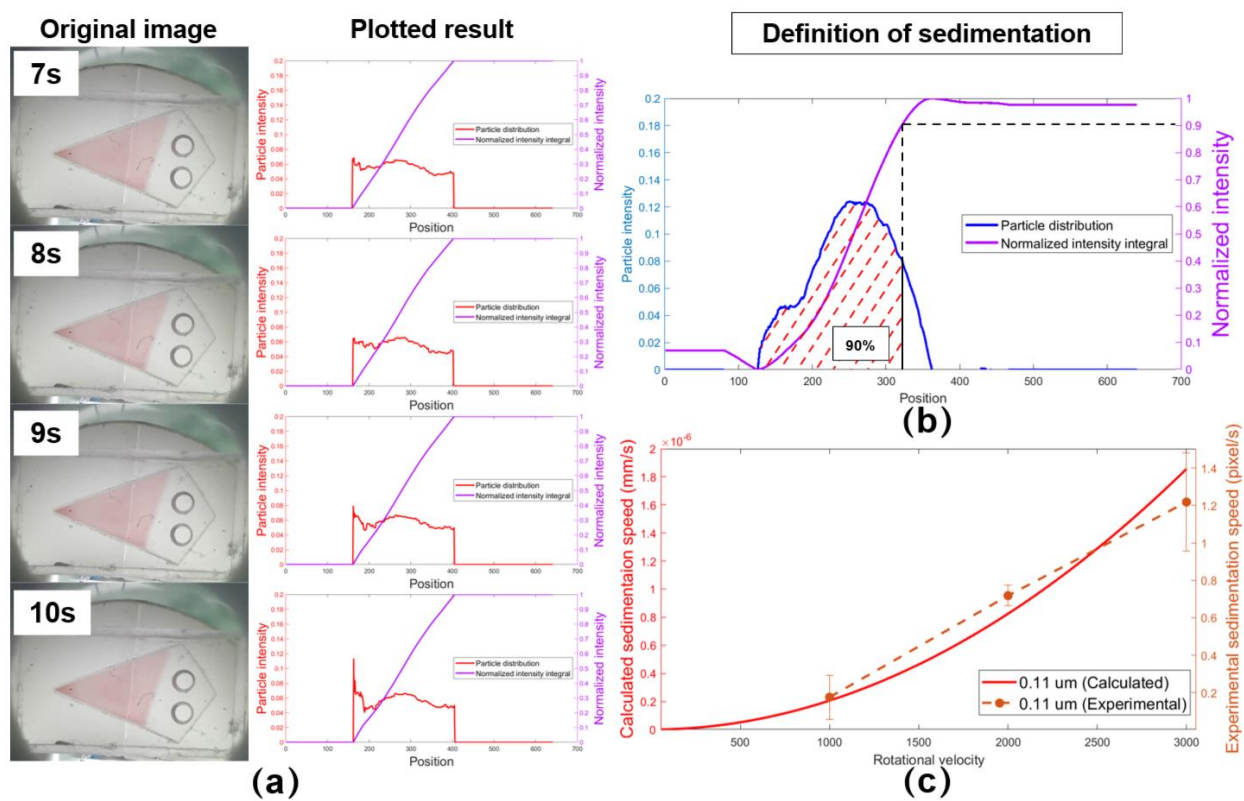

**Fig. 3 Experimental results of monitored sedimentation process by LIAT spectrophotometer.** (a) Sequential frames for the sedimentation process of 0.11 μm PS beads captured by LIAT set-up and corresponding plotted particle distribution. With sedimentation duration increases, the target particle intensity of the chip's left side increases. (b) Definition of sedimentation by an integral function of particle intensity. The x-position for normalized integral intensity equals 0.9 correlates to the position of 90% particle sedimented. Thus the sedimentation speed is determined by calculating the moving speed of x-position for normalized integral intensity equals 0.9. (c)

Comparison between experimental result by above technique and calculated result by theoretical calculation.

We also demonstrated the field portability and on-site operation of our LIAT imaging spectrophotometer by performing an automated hematocrit test (Hct) of whole blood from finger-prick volumes. Figure 4 highlights the performance of this automated image alignment and HSV-based segmentation process, and the image quality achieved by the device, showcasing separated blood images at different stages with both their initial captured raw images and the segmented images of different components. As the centrifugation process carried on, the area of plasma increased and the area of red blood cells (RBCs) decreased, with a stable area ratio of 0.575 (variation < 0.05%). This result showed good agreement with the analysis performed by clinical Hct test, while with an ultra-low sample volume. In clinics, the Hct method requires venipuncture. The analysis process is based on the microhematocrit centrifuge, which takes more than \$500 per facility. The percentage analysis of RBCs in blood is based on measuring the length of the RBCs column and whole blood column utilizing a hematocrit reader or any ruled apparatus. Whereas our analysis is based on imaging of the finger-prick volume blood sample itself during an all-in-one centrifugation process within common lab centrifuges automatically. Differences in sample collection, imaging, and data processing techniques might cause some systematic differences between the two analysis composition metrics reported in Fig. 4 and Fig. 5. However, both methods are self-consistent, and therefore the relative differences that are observed in blood composition are comparable, illustrating good agreement between our results and the analysis performed by traditional clinical methods.

For on-site experiments, the one-stop operation includes real-time image capturing, image analysis, and programmable interruption. After setting the interruption condition, the blood separation process and plotting of different components are real-time displayed by the developed GUI. Here we set it as Area $_{RBCs}$ /Area $_{whole\ blood}$ decreased from 1 to less than 0.6 with a variation of less than 0.05% for constant 10 seconds. Once the interruption command is triggered, it's recognized as "blood separation complete", and the centrifugation process is automatically stopped (relay connected to centrifuge power cable). In the centrifugation process, the frames of the first 10 seconds were collected for finding background and pre-saving segmentation parameters. The bubbles within the blood sample were also excluded during this process. After that, the collected images were real-time analyzed and displayed. The whole separation process takes around 5 minutes, whereas the "blind" conventional clinical blood separation takes 15 minutes. Moreover, our device provides continuous direct visualization of the whole process, automatic control, less time cost, less sample cost, higher precision and higher RBCS cell viability. These results demonstrate the capability of our portable imaging spectrophotometer for Point-of-care testing (POCT) adapting to common lab centrifuges.

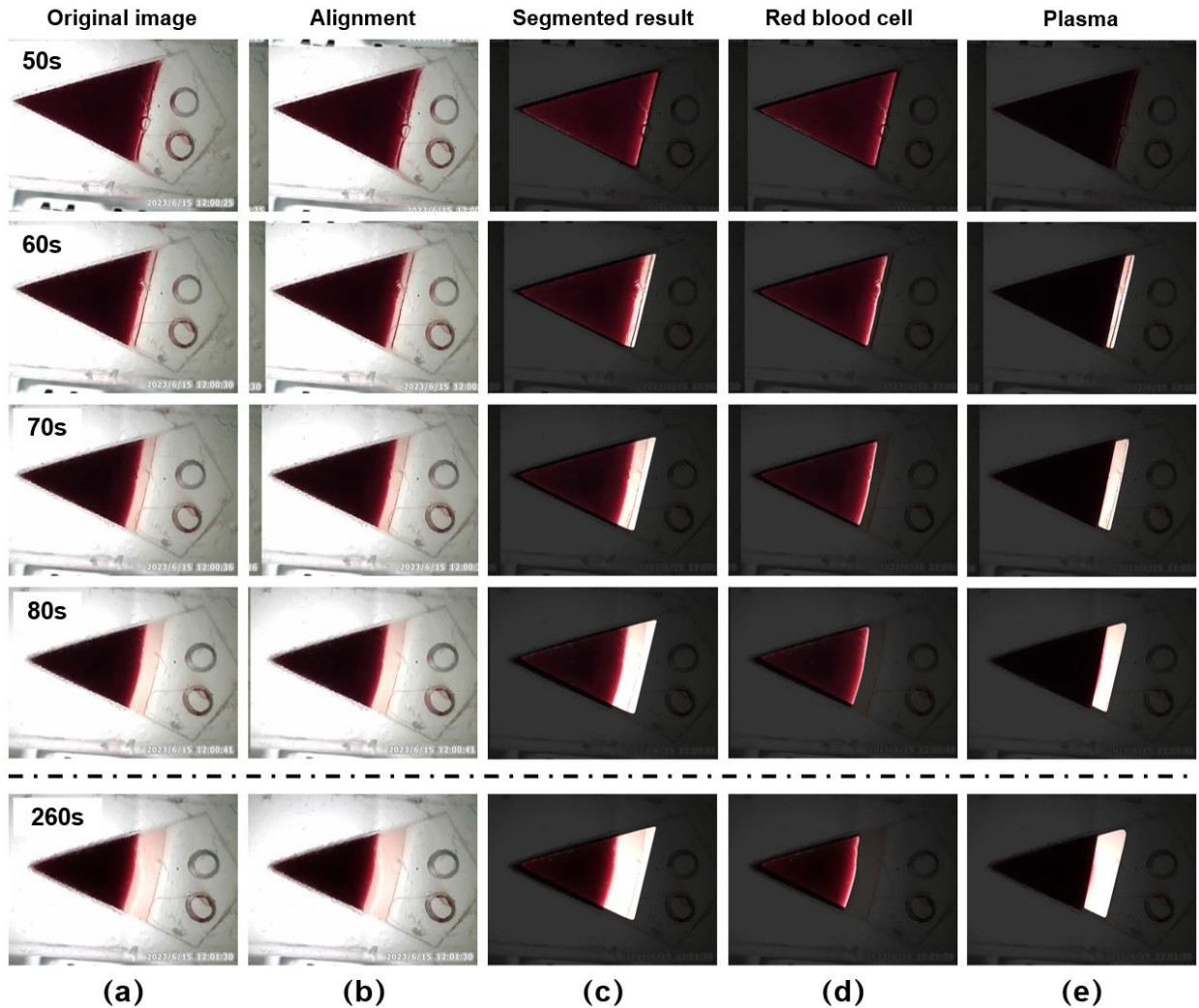

**Fig. 4 Image analysis results of representative frames for blood separation by LIAT spectrophotometer.** From the first line to the penultimate line, frames were presented every ten seconds. For the last line, the blood separation is complete. (a) Sequential frames for blood separation captured by LIAT spectrophotometer. The captured time of each frame is labeled in the upper left corner. Here the original images were only applied crop operation for better presentation. (b) Captured frames were applied alignment operation for precise analysis. (c) Segmentation result of whole blood sample region. (d) and (e) Red blood cells region and plasma region were segmented respectively based on gray value for determining region areas.

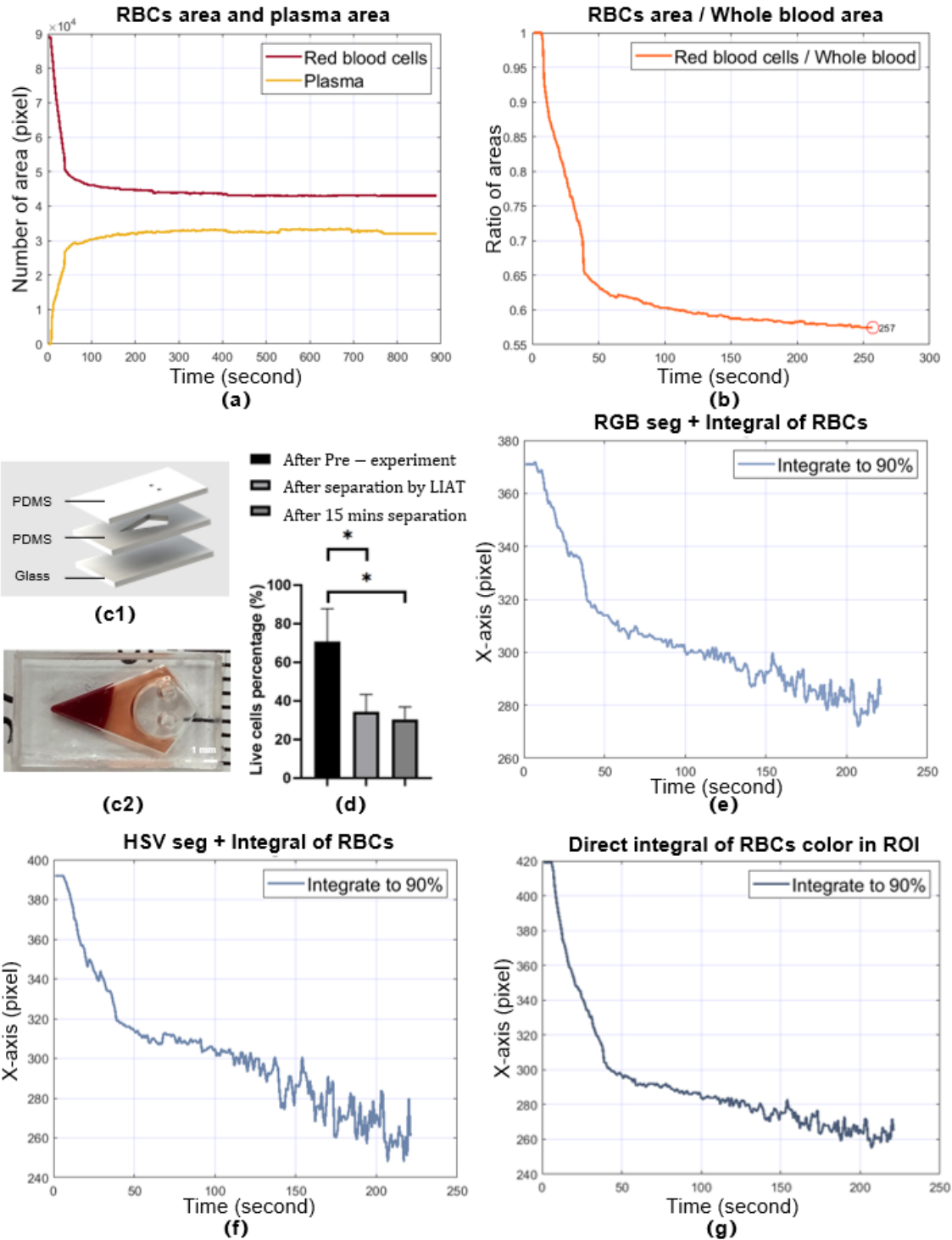

**Fig. 5 Field test results of programmable interrupt blood separation by LIAT imaging spectrophotometer. Comparison of different data analysis method.** (a) In the beginning, all the sample region is recognized as a red blood cells (RBCs). The area of plasma is 0. With the blood sample being separated, the area of RBCs decreased and the area of plasma increased. After 5 minutes, the spectral lines of different components become smooth. (b) The on-site experiment

result of programmable interruption blood separation. Plotted result of area ratio (area of RBCs versus area of whole blood) based on definition of hematocrit test (Hct). When the experiment run for 257 seconds, the interruption function was triggered (Area $_{RBCs}$/Area $_{whole\ blood}$ decreased from 1 to less than 0.6 with a variation of s than 0.05% for continuous 10 frames). The segmentation was applied using the HSV-based method. (c1) Design of portable microfluidic chip for LIAT imaging spectrophotometer set-up. (c2) Image of blood sample in the microfluidic chip after blood separation is complete. (d) RBCs viability of different experiment stages. The pre-experiment including start and stop takes around 1 minute. The whole process of automated blood separation with data analysis by LIAT imaging spectrophotometer takes 5 mins. In clinics, the blood separation takes 15 mins. Utilizing our device, blood separation and analysis can be realized in an automated manner with less time cost and higher cell viability compared to conventional methods. (e) Data analysis result by RGB-based segmentation and integral calculation of segmented RBCs area. A lot of noise was induced when the blood separation is complete, as a result of the color of RBCs getting darker due to its concentration. (f) Data analysis result by HSV-based segmentation and integral calculation of segmented RBCs area. Noise is still significant. (e) Data analysis result by direct integral calculation of RBCs color in (Region of interest) ROI area without segmentation of different components. Noise is still significant.

## 3. Discussion and conclusion

The throughput and accuracy of the LIAT imaging spectrophotometer that collects colored data over a two-dimensional area is determined by several factors, but most importantly it is governed by the acquired image quality, color calibration, and light source. We designed our portable LIAT imaging spectrophotometer to achieve the highest resolution allowed by the pixel size of the image sensor, which is 1632 columns by 1232 rows (2,010,624 pixels). We calibrated our color filtering by different methods for different tasks. For the single color task (sedimentation speed determination), we extracted the RGB of the target color using Matlab and fine-tuned the RGB setting based on the detection results of collected images. For multiple-color tasks (PS beads separation), we extracted and fine-tuned the RGB of target colors respectively from separated images with the same method. We optimized the accuracy of the detection results utilizing normalizing, enhancing contrast, and de-nosing by filtering and morphological operations. We optimized the accuracy of the detection results by applying the color deconvolution method. For the images with components of high contrast (blood separation task), we optimized the image analysis method by extracting the HSV (Hue, Saturation, Value) information of different components. Moreover, we also applied noise reduction and image alignment for optimization. The application of the proposed platform can be promoted even more by replacing the light source from white light LED with specific wavelength light sources (e.g., RGB pulsed illumination light sources).

In addition to the aforementioned parameters, the processing speed of the controlling laptop that triggers the termination through closed-loop control can also be a limiting factor. Currently, the device can be run in two modes based on the requests: offline mode and real-time mode. For offline modes, we collect and save the original images using a LIAT imaging spectrophotometer. Later after the experiment has been complete, we can perform image analysis by reading the images from folders. This is also the approach for precise fine-tuning the parameters (e.g., threshold setting) for optimization. Second, the device is capable of monitoring the centrifugation process and actively performing real-time on-demand termination. Regarding the runtime analysis, at present, the Wi-Fi image transmission, image analysis, and triggering relay procedure take ~1.735 s for each full FOV frame, in which up to 3 target colors can be real-time plotted with parallel computing. The major computational operations are (1) channel region segmentation from the full FOV image (8 bit, $1024 \times 1024$ pixels $\times$ 3 color channels) by finding background and binary masking (~1.198 s per image), (2) immediate color deconvolution and plotting (~0.720 s), and saving of the images on an internal solid-state drive (~ 0.668 s per image). (3) integral calculation if necessary (~0.689 s for each integral calculation). The Wi-Fi image transmission from LIAT inside the centrifuge to the laptop takes less than 0.060 s per image (Supplementary Fig. 8). In real-time mode, we capture and display 1 frame per second. The programmable interruption operation by relay takes ~5 s. In summary, the auto-response behavior in our LIAT system can happen within a delay of less than 7 s.

In our experiments, since the background signal has negligible spatial variation and is removed from FOV via digital background removal, we can tolerate a certain loss of light transmission due to potential fouling. Furthermore, since the microfluidic chips of our imaging spectrophotometer are disposable and can be easily replaced, a routine evaluation of the background signal can be applied to automatically alert users of the need for replacement.

To clarify the limiting use cases, we have identified that our current prototype provides stable performance in centrifugation with rotational velocity up to 6,000 r.p.m., while minor deformation of the 3D-printed holder has been observed at rotational velocities beyond, which can be resolved by replacing 3D-printed material from resin to Nylon[18]. As our current prototype is a field-portable imaging spectrophotometer, the time of endurance is more than 4 hours with easy replacement of the Li-ion battery. This prototype can operate at a physical distance of up to 100 meters from the controlling laptop by simply utilizing the Wi-Fi camera for long-range communication. Owing to its low hardware complexity in comparison with other imaging spectrophotometer technologies, the component cost of the system is less than $25, and 4 sets of systems can be operated within one centrifuge simultaneously. This remarkable cost-effectiveness opens up exciting opportunities for centrifugation microbiology research and paves the way for the creation of a network of computational imaging spectrometers at an affordable price for large-scale and continuous monitoring of centrifugal processes in general.

## 4. Materials and methods
### 4.1 Lab-in-a-Tube imaging spectrophotometer setup

The LIAT imaging spectrophotometer system is a multi-component system comprising a Wi-Fi camera, a microfluidic chip, a lithium battery, a white light LED, and a 3D-printed cartridge. The Wi-Fi camera was designed using Altium Designer software and fabricated through outsourcing (Supplementary Fig. 4). It features an OV2640 CMOS sensor with an optical size of 1/4″ and a resolution of 1632×1232 pixels at 15 FPS (frame per second) for imaging. The cartridge, with a diameter of 21 mm and a length of 65 mm, was designed using Solidworks2016 and 3D printed using resin, making it suitable for use in a common centrifuge system (Supplementary Fig. 5). The choice of electronic components, and soldering schemes, and orientation of various items have been optimized to minimize mechanical deformation associated with extreme centrifugal force. All components were assembled seamlessly and tested before conducting experiments, ensuring the proper functioning of the system. A USB relay was connected between the laptop and the power cable of the centrifuge, allowing for closed-loop control of the system using Matlab. The design files and control codes can be found in the Supporting Information.

### 4.2 Image analysis
The video capturing setting was 1632×1232 pixels. Before image analysis, the video was snapped into image frames. The frames were cropped into 640×480 pixels to focus on the sample region. For automatic color analysis of the monitoring images collected in the continuous centrifugation process (see Fig. 2), the brightness was normalized and the background was eliminated first. This is achieved by calculating a time-averaged image of the preceding 20 images containing only the static objects and subtracting it from the present raw image. This yields a background-subtracted full FOV image in which only the objects newly introduced by the flow are present. These objects are automatically detected and segmented from the full FOV for individual processing (see Supplementary Figure S2). The full FOV background-subtracted image is first Gaussian-filtered and converted into a binary image by hard thresholding. The binary contours with an area of a few pixels are removed to reduce misdetection events due to sensor noise. The resulting binary contours represent the shapes and locations of the objects appearing in the FOV, and their morphological information is used to filter each contour by certain desired criteria (e.g., color distribution). The coordinate of inlets of the filtered contour is used to segment its corresponding image (parts right beside inlets are neglected). After segmentation, the intensity of target colors was detected and plotted respectively. We should emphasize that not only is it feasible to extract all objects in the FOV but it is also possible to perform integral, differential, and other operations of the objects of interest for a specific goal by our approach. Thus, we can better utilize the computational resources of the laptop and maintain multi-functional real-time processing for different samples.

### 4.3 Graphical user interface
We developed a GUI to operate the device. Through this GUI, the cut-off trigger condition can be specified, such as the ratio of different components. The GUI gives a display of captured original image, segmented image, and corresponding plotting of target colors in a real-time manner, thus allowing visual inspection during centrifugation processes with and without background subtraction. The user can specify whether to digitally save the raw images, background-subtracted images, plot results, and calculated values. The GUI can also be run in demo mode to analyze pre-saved image datasets without the presence of the imaging Lab-in-a-Tube.

### 4.4 Microfluidic chips design and Fabrication
The molds of microfluidic chips were designed using SolidWorks 2016 and manufactured using a CNC (computer numerical control) machining machine (YB4030X, Guangzhou Yubang Machinery Co Ltd., Guangzhou, China). Subsequently, PDMS prepolymers (Sylgard 184, Dow, Inc., Korea) with a base-to-curing ratio of 10:1 was mixed, degassed, and poured onto the molds before being baked at 60 °C for 10 hours. After the PDMS was cured, it was peeled off and cut into desired shapes, followed by punching out the inlets and outlets. The PDMS slabs and glass slides were treated with oxygen plasma for 1 minute, placed in contact, and baked at 110°C briefly for bonding. The devices were then baked at 55°C for 24 hours to turn the channel surface hydrophobic. The fabricated chip has been proven to resist a rotating speed of 6,000 r.p.m. with no leakage observed.

### 4.5 Sample Preparation and Analysis

Fingertip blood was collected from healthy consenting volunteer into 3 mL Blood Collection Tubes (VACUETTE®, Greiner Bio-One International GmbH, Sweden) and used within 4 hours of donation. Human After the sedimentation process, the RBCs cells were extracted from a microfluidic chip and suspended in 1 ml PBS (Phosphate-buffered saline) buffer. Afterward, the suspended cells were mixed with 0.4% trypan blue by 1:1. The cell number and cell viability were obtained using an Automated Cell Counter (TC20™, Bio-rad).

### 4.6 Statistical analysis

In all cases, data represent mean ± SD with n ≥ 3. A two-sided two-sample t-test was adopted for hypothesis testing, and significance was defined as $p \leq 0.05$. GraphPad software was used to analyze all data for statistical significance.

### 4.7 Numerical simulation

The numerical simulation of the single particle in the Results section was studied by Mathematica®.

To understand the dynamics of the single re-circulating flow, here we subject a single particle within the microfluidics (in a rotating reference frame). We investigated the dynamical equilibrium of the single particle as followed:

$$\vec{F_b} - \vec{F_d} - \vec{F_{cor}} = m\ddot{x} \tag{1}$$

Here $\vec{F_b}$ is the buoyancy induced by the centrifugal force; $\vec{F_d}$ is the drag force; and $\vec{F_{cor}}$ is the Coriolis force. $m$ is the mass of a particle and $\ddot{x}$ is the coordinates of the particle. Substituting the expressions for each component, we have:

$$m\ddot{x} = \Delta\rho V \omega^2 x - 6\pi\mu r_c \dot{x} - 2m\omega \times \dot{x} \tag{2}$$

where $\Delta\rho = \rho_b - \rho_f$ is the difference between the PS beads' density $\rho_b$ and fluid density $\rho_f$. $\mu$ is the viscosity of the fluid, and $r_c$ is the particle radius.

Here we define the characteristic parameters as followed:

$$\omega_0 = \frac{3\pi\mu r_c}{m} = \frac{9\mu}{4\rho_b r_c^2} \tag{3}$$

$$\alpha = \frac{\Delta\rho}{\rho_b} \tag{4}$$

$$\beta = \frac{\omega}{\omega_0} \tag{5}$$

The characteristic frequency is determined by the mass, radius, and friction constant. Here $\alpha$ and $\beta$ are dimensionless.

To simplify the derivation, we apply $\omega_0$ as the natural time unit:

$$t = \frac{\tau}{\omega_0} \tag{6}$$

By taking the derivation, the equation (2) is simplified of dimensionless as followed:

$$x'' + 2x' + 2\frac{\omega}{\omega_0} \times x' - \frac{\Delta\rho}{\rho_b}\frac{\omega^2}{\omega_0^2}x = 0 \tag{7}$$

Or in terms of matrix

$$X'' + 2(1 - \beta i\sigma_y)X' - \alpha\beta^2 X = 0 \tag{8}$$

After simplifying the matrix and solving the procedure (which can be found in Supplementary Information), we got the general solution:

$$X(\tau) = \left(Ae^{\left(-1+i\beta+\sqrt{(1-i\beta)^2+\alpha\beta^2}\right)\tau} + ae^{\left(-1+i\beta+\sqrt{(1-i\beta)^2+\alpha\beta^2}\right)\tau}\right)\binom{1}{i}$$
$$+ \left(Be^{\left(-1-i\beta+\sqrt{(1+i\beta)^2+\alpha\beta^2}\right)\tau} + be^{\left(-1-i\beta+\sqrt{(1+i\beta)^2+\alpha\beta^2}\right)\tau}\right)\binom{1}{-i} \tag{9}$$

The $a$ and $b$ term decay exponentially. For simplification, we neglect them when the time is large enough.

Thus, the solution is simplified as:

$$X(\tau) = \left(Ae^{\left(-1+i\beta+\sqrt{(1-i\beta)^2+\alpha\beta^2}\right)\tau}\right)\binom{1}{i} + \left(Be^{\left(-1-i\beta+\sqrt{(1+i\beta)^2+\alpha\beta^2}\right)\tau}\right)\binom{1}{-i} \tag{10}$$

Let's define

$$\sqrt{(1+i\beta)^2 + \alpha\beta^2} = Me^{i\phi} = R + iI \tag{11}$$

$$\sqrt{(i\beta)^2 + \alpha\beta^2} = Me^{-i\phi} = R - iI \tag{12}$$

Then

$$X(\tau) = e^{(R-1)\tau}\left[Ae^{-i(I-\beta)\tau}\binom{1}{i} + Be^{i(I-\beta)\tau}\binom{1}{-i}\right]$$

$$= e^{(R-1)\tau} e^{\Omega(I-\beta)\tau} X(0) \tag{13}$$

Here $\Omega$ is the matrix defined above.

In polar coordinates, the above equation indicates the particle orbits the center of rotation with uniform velocity:

$$\theta(\tau) = \theta(0) + (I - \beta)\tau \tag{14}$$

$$r(\tau) = r(0)e^{(R-1)\tau} \tag{15}$$

In the original unit, the orbiting angular velocity is:

$$\Omega = \Im\sqrt{(\omega_0 + i\omega)^2 + \frac{\Delta\rho}{\rho_b}\omega^2} - \omega \tag{16}$$

$$\gamma = \Re\sqrt{(\omega_0 + i\omega)^2 + \frac{\Delta\rho}{\rho_b}\omega^2} - \omega_0 \tag{17}$$

Plot the $\Omega$ and $\gamma$ as shown in Fig. 2(c1) and (c2), we find we always have $\Omega < 0$ and $\gamma > 0$. $\Omega < 0$ indicates the particle tends to rotate in the opposite direction to the rotating frame. $\gamma > 0$ indicates the particle moving away from the rotational center axis. With radius and rotational angular velocity increase, the absolute value of $\Omega$ and $\gamma$ increase monotonically. This conclusion has also been proven by experiments: sedimentation and vortex exist simultaneously.

To quantify the significance of vortex, we compared the two velocities by the following definition:

$$\Lambda = \frac{\Omega}{\gamma} \tag{18}$$

Roughly this quantifies the degree that the vortex rotates before sedimentation. Supplementary Fig. 1 shows $\Lambda$ also increases monotonically with the parameter *decreases* but decrease with $\frac{\Delta\rho}{\rho_b}$. And $\Lambda$ increases particularly fast when the density difference is far smaller than the density of the particle.

The numerical simulation of fluid was studied by a simulation model established by COMSOL Multiphysics® (v4.3, COMSOL Incorporation). Single-phase, steady-state simulations were performed under laminar flow conditions. The model is straightforward to set up using a Laminar Flow interface. The unphysical boundary condition was used in this model. The flow within the microfluidic channel is manipulated by angular velocity-dependent centrifugal force (along the x-axis) and Coriolis force (along the y-axis). The system is simplified from 3-dimensional to 2-dimensional as its symmetry is along the z-axis. For the rounded chip applied in the LIAT Centrifuge system, turbulence has been observed in the simulation result of velocity magnitude (Fig. 6). This correlates with the vortex phenomenon observed by our system. The pressure increased with the x-coordinate increase, which correlates with the sedimentation phenomenon observed by our system.

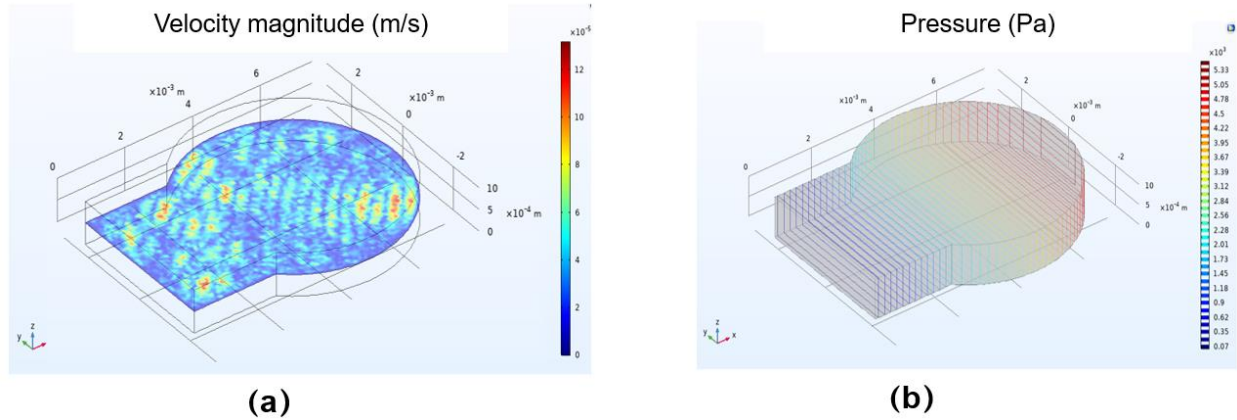

**Fig. 6 Numerical simulation of centrifugal fluidity in the LIAT system.** (a) Velocity magnitude simulation result of fluidity in the LIAT Centrifuge system. (b) Simulation result of a pressure simulation result of fluidity in the LIAT Centrifuge system. The pressure increased with the x-coordinate increase.

Sedimentation on a microfluidic device is governed by Stokes' Law, and when applied to centrifugal microfluidics like our system, the settling velocity ($V$) of a particle can be calculated as:

$$V = \frac{2R^2\omega^2 r}{9\eta_0} \times (\rho_p - \rho_0) \tag{19}$$

where $R$ is the Stokes' radius, $\omega$ is the angular velocity (Hz), $r$ is the average radial position, $\eta_0$ is the viscosity, $\rho_p$ is the particle density, and $\rho_0$ is the fluid density.

## 6. Acknowledgements

The authors are grateful to the funding support from the Hong Kong Research Grants Council (GRF14216222, GRF14203821, GRF14204621, GRF14207920, ECS24211020, CUHK 14207121 and GRF14207419), the Innovation and Technology Fund of Hong Kong (ITS/240/21), Shenzhen-Hong Kong-Macau Science and Technology Program (Category C) of Shenzhen Science, Technology and Innovation Commission (SGDX20220530111005039).

The authors would like to acknowledge Dr. Shiyue Liu, Dr. Guangyao Cheng, Dr. Shutian Zhao, Dr. Md Habibur Rahman, Ms. Yijin Wang, Ms. Shaojun Liu, Mr. Nelson So, Mr. Yujie Su, Mr. Zhaoyu Liu, Mr. Sai Mu Dalike Abaxi, Ms. Syeda Aimen Abbasi (Department of Biomedical Engineering, The Chinese University of Hong Kong), Mr. Zheyuan Zhang (Department of Biomedical Engineering, Northwestern University), Dr. Mingzu Liu (Department of Physics, Penn State University), Ms. Yuhan Ju (Department of Mechanical Engineering, National University of Singapore) and M5Stack Technology Co., Ltd for their support in the project development.

## 7. Data Availability Statement

The data that support the findings of this study are available from the corresponding author upon reasonable request.

## 8. Author information

Authors and Affiliations

**Department of Biomedical Engineering, The Chinese University of Hong Kong, Shatin, Hong Kong SAR, 999077, China**

Yuanyuan Wei, Dehua Hu, Tsz Kin Chan, Yi-Ping Ho, Wu Yuan & Ho-Pui Ho

**Electrical and Computer Engineering Department, University of California, Los Angeles, CA, 90095, USA**

Bijie Bai

**Department of Physics, The Chinese University of Hong Kong, Shatin, Hong Kong SAR, 999077, China**

Chenqi Meng

**Department of Orthopaedics and Traumatology, Li Ka Shing Institute of Health Sciences, The Chinese University of Hong Kong, Hong Kong SAR, 999077, China**

Xing Zhao

**Bionic Sensing and Intelligence Center, Institute of Biomedical and Health Engineering, Shenzhen Institutes of Advanced Technology, Chinese Academy of Sciences, Shenzhen, 518055, China**

Yuye Wang

**Centre for Biomaterials, The Chinese University of Hong Kong, Hong Kong SAR, 999077, China**

Yi-Ping Ho

**Hong Kong Branch of CAS Center for Excellence in Animal Evolution and Genetics, Hong Kong SAR, 999077, China**

Yi-Ping Ho

**The Ministry of Education Key Laboratory of Regeneration Medicine, Hong Kong SAR, 999077, China**

Yi-Ping Ho

Contributions

Y. Wei developed the Lab-in-a-Tube prototype and relay-based closed-loop control of the centrifuge. B. Bai developed the image analysis algorithm. C. Meng developed the theory of a single particle in a rotating reference system. Y. Wei realized the Matlab codes for real-time monitoring of the sedimentation process. D. Hu realized the Matlab codes for real-time blood separation analysis. Y. Wei and T. Chan realized the GUI. Y. Wei developed the centrifugal microfluidic simulation. Y. Wei and T. Chan performed the experiments. Y. Wei, D. Hu, and T. Chan performed data analysis. X. Zhao and Y. Wei designed the PCB. Y. Wei wrote the manuscript with contributions from all authors. H. Ho conceived the project and supervised the research.

Corresponding author

Correspondence to Ho-Pui Ho.

9. Ethics declarations

Conflict of interest

H.H. and Y.W. have a pending patent application on the presented Lab-in-a-Tube (LIAT) imaging spectrophotometer.

10. Additional information

Electronic supplementary material

Supplementary Information